\documentclass[cameraready]{Interspeech}

\title{Universal Speech Content Factorization}

\author[affiliation={2}, orcid=0009-0005-9601-4515]{Henry Li}{Xinyuan}
\author[affiliation={2}, orcid=0009-0007-2658-2220]{Zexin}{Cai}
\author[affiliation={2}, orcid=0000-0001-7826-2850]{Lin}{Zhang}
\author[affiliation={1}, orcid=0000-0002-7449-5726]{Leibny Paola}{Garc\'ia-Perera}
\author[affiliation={1}, orcid=0000-0001-8078-3305]~{Berrak~Sisman}{}
\author[affiliation={1}, orcid=0000-0001-5976-0897]{Sanjeev}{Khudanpur}
\author[affiliation={2}, orcid=0000-0002-6097-9164]{Nicholas}{Andrews}
\author[affiliation={2}, orcid=0000-0002-5423-7754]{Matthew}{Wiesner}

\address{
    $^1$ Center for Language and Speech Processing, Johns Hopkins University, USA \\
    $^2$ Human Language Technology Center of Excellence (COE), Johns Hopkins University, USA
}

\email{xli257@jhu.edu}

\usepackage{comment}
\usepackage{cite}
\usepackage{url}
\usepackage[table]{xcolor}
\usepackage{colortbl}
\usepackage{pgf}
\usepackage{xparse}

\RequirePackage{etoolbox}
\RequirePackage{pgf} 
\RequirePackage{xcolor} 
\definecolor{high}{HTML}{ef3b2c}  
\definecolor{low}{HTML}{fff7bc}  

\newcommand{\gradientcell}[6]{
    \ifdimcomp{#1pt}{>}{#3 pt}{#1}{
    \ifdimcomp{#1pt}{<}{#2 pt}{#1}{
         \pgfmathparse{int(round(100*((#1-#2)/(#3-#2))))}
        \xdef\tempa{\pgfmathresult}
        \cellcolor{#5!\tempa!#4!#6} #1
    }}
 }

\newcommand{\reversegradientcell}[6]{
    \ifdimcomp{#1pt}{>}{#3 pt}{#1}{
    \ifdimcomp{#1pt}{<}{#2 pt}{#1}{
         \pgfmathparse{int(round(100*((#3-#1)/(#3-#2))))}
        \xdef\tempa{\pgfmathresult}
        \cellcolor{#5!\tempa!#4!#6} #1
    }}
 }

\begin{document}

\maketitle

\begin{abstract}

We propose Universal Speech Content Factorization (USCF), a simple and invertible linear method for extracting a low-rank speech representation in which speaker timbre is suppressed while phonetic content is preserved. USCF extends Speech Content Factorization, a closed-set voice conversion (VC) method, to an open-set setting by learning a universal speech-to-content mapping via least-squares optimization and deriving speaker-specific transformations from only a few seconds of target speech. We show through embedding analysis that USCF effectively removes speaker-dependent variation. As a zero-shot VC system, USCF achieves competitive intelligibility, naturalness, and speaker similarity compared to methods that require substantially more target-speaker data or additional neural training. Finally, we demonstrate that as a training-efficient timbre-disentangled speech feature, USCF features can serve as the acoustic representation for training timbre-prompted text-to-speech models. Speech samples and code are publicly available.\footnote{Code release: \scriptsize{\url{https://github.com/HSTEHSTEHSTE/uscf}}; \footnotesize Speech samples: \scriptsize{\url{https://hstehstehste.github.io/Projects/uscf.github.io/index.html}}}


\end{abstract}

\section{Introduction}
Recent self-supervised learning (SSL) models for speech, such as WavLM, exhibit pronounced geometric structure in their feature spaces, with empirical analyses demonstrate that phonetic content dominates feature variance, and that frames corresponding to the same phoneme forming tight clusters across speakers \cite{wavlm_phoneme_cluster, wavlm_phoneme_recognition}. This finding has significant consequences for the downstream task of voice conversion (VC), where the typical goal is to modify the speaker identity while preserving the linguistic content, enabling a class of training-free voice conversion methods that operate directly in SSL space \cite{knnvc, linearvc}.

Notably, the voice conversion system kNN-VC \cite{knnvc} showed that VC can be performed by replacing each WavLM feature frame extracted from an input utterance with its nearest neighbor from a target speaker's WavLM feature collection. The success of kNN-VC suggests a structural property of the WavLM feature space: for each phoneme, frames from different speakers reside within a consistent subspace. This hypothesis is supported further by LinearVC \cite{linearvc}, which showed that a content-preserving approximate linear projection can be found between the WavLM features of two speakers. Building on this observation, Speech Content Factorization (SCF) \cite{linearvc} proposed projecting WavLM features into a shared low-rank representation encoding phonetic content, and reconstructing speaker-specific features through learned linear transformations, thereby enabling high-quality VC without additional model training.

\begin{figure}[t!b]
  \centering
  \includegraphics[width=\linewidth]{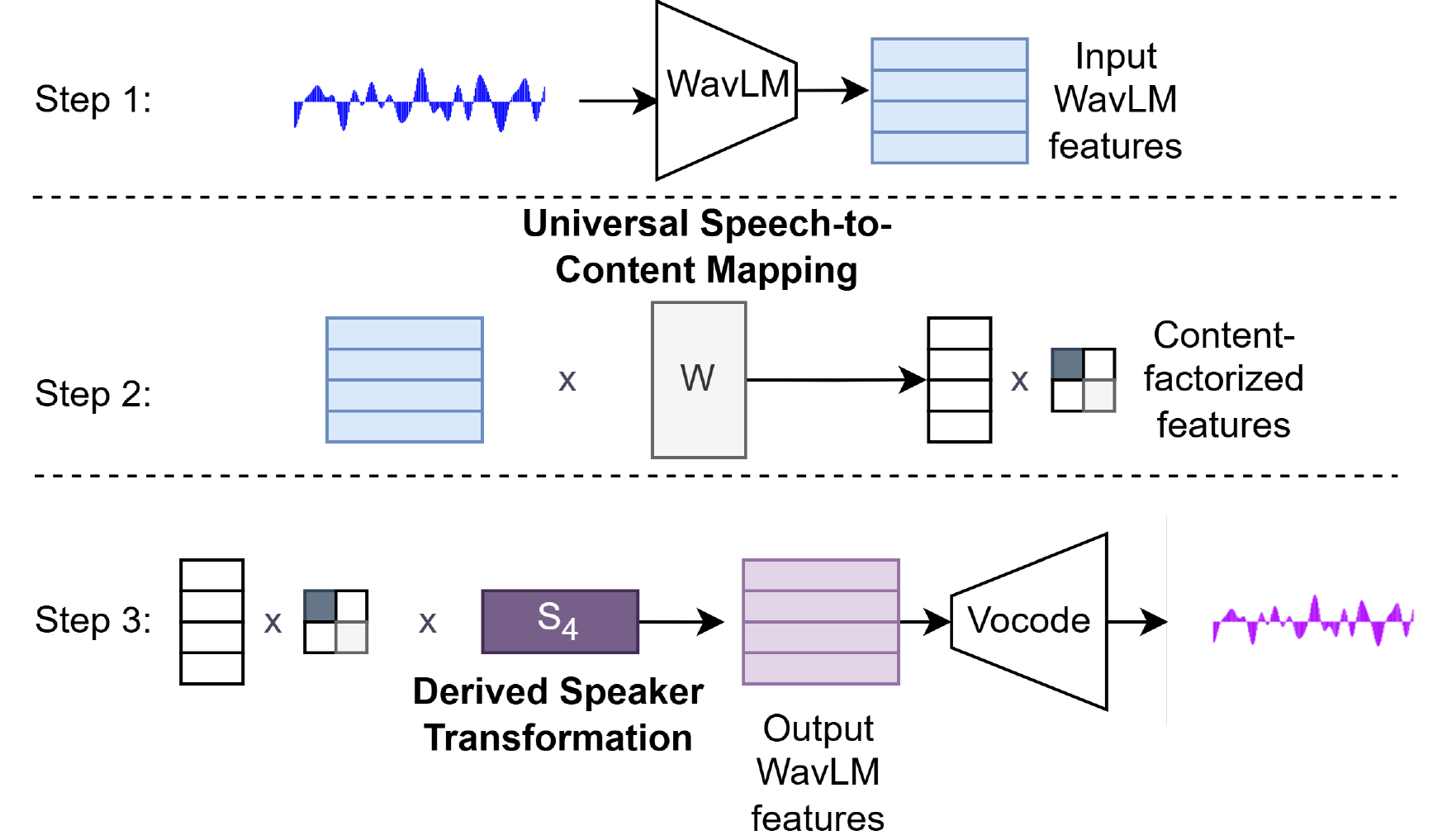}
  \caption{Full pipeline for voice conversion using USCF.}
  \label{fig:s}
\end{figure}

\begin{figure}[t!b]
  \centering
  \includegraphics[width=\linewidth]{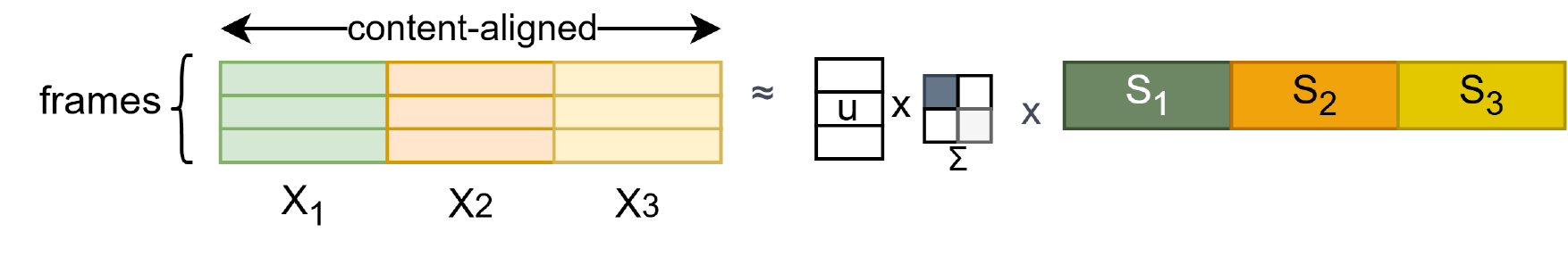}
  \caption{Decomposing speech into a content-factorized form through SCF. $\mathbf{X}_i$ are content-aligned WavLM features for different speakers. Content alignment for $\mathbf{X}_i$ is performed through kNN matching.} 
  \label{fig:scf}
\end{figure}

\begin{figure}[t!b]
  \centering
  \includegraphics[width=\linewidth]{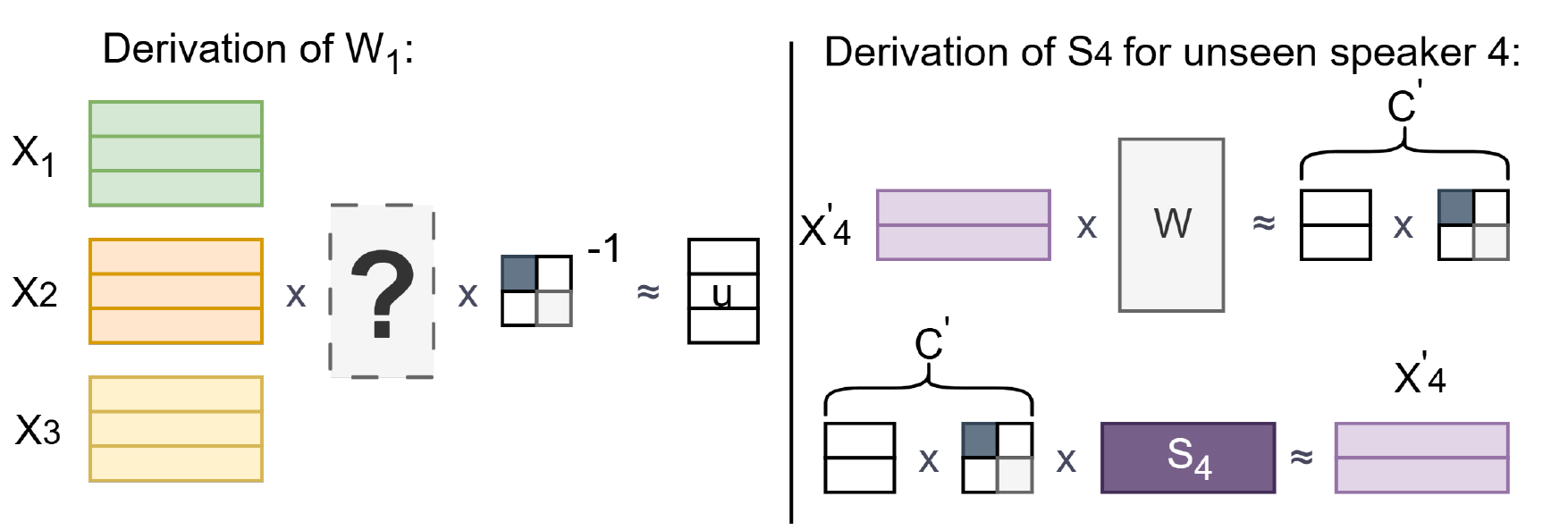}
  \caption{Left: formulation for $\mathbf{W}_1$, one of our proposed universal speech-to-content mappings. Right: Derivation of speaker transformation matrix $\mathbf{S}_4$ for unseen speaker $4$}
  \label{fig:ws}
\end{figure}




Despite its simplicity and effectiveness, SCF is a closed-set method: extracting a content-factorized representation from a speaker requires that the speaker be included in the set of speakers used to derive the factorization. This restriction limits its applicability in downstream scenarios such as open-set VC or timbre-prompted TTS, where unseen speakers must be supported without recomputing the decomposition. For example, in order to train TTS models on crowd-sourced or web-crawled datasets with diverse speaking styles such as CommonVoice \cite{commonvoice} or Emilia \cite{emilia}, training SCF on all of the speakers present would both be prohibitively expensive, and exclude many speaker that don't have enough speech present.

To address this limitation, we propose \textbf{Universal Speech Content Factorization (USCF)}, an open-set extension of SCF that enables speaker-agnostic content extraction and one-shot speaker adaptation. 
Leveraging the linear structure of SCF, we derive a speaker-agnostic universal speech-to-content mapping via least-squares optimization, and infer speaker-specific content-to-speech transformations from as little as a single target-speaker utterance using linear estimation. Our system achieves competitive performance compared to SSL-structure-based baselines and to SCF. Our contributions are as follows:
\begin{itemize}
    \item We propose USCF, a universal speech-to-content mapping, by showing that the linear structure underlying SCF generalizes to unseen speakers. We show that a universal speech-to-content mapping can be computed using a simple least-square formulation, while speaker-specific content-to-speech transformations can be estimated from only a few seconds of target speech.
    \item We evaluate USCF as a zero-shot VC system and show competitive performance in intelligibility, naturalness, and speaker similarity compared to baselines.
    \item We further show that the USCF representation can serve as an alternative acoustic target for text-to-speech (TTS) systems.
    \item We perform embedding analyses on USCF representations and show that they contain less speaker information than other speaker-factorized representations while effectively preserving speech content.
\end{itemize}

\subsection{Related Works: Speech Disentanglement}

In contrast to SSL-space methods, most speech disentanglement approaches rely on explicitly trained generative models. A common strategy is to use a Variational Autoencoder (VAE) \cite{vae} that reconstructs speech from a bandwidth-limited intermediate representation encoding one factor, together with an auxiliary encoder that provides the disentangled factor \cite{continuous_codec, speechsplit2, adv_disentangled, contentvec, dvae, naturalspeech3}. Such methods have been applied to VC \cite{disentangle_for_vc, genvc, sisman2020overview}, expressive speech translation \cite{disentangle_for_st}, speech anonymization \cite{disentangle_for_privacy, npu_ntu}, and accent conversion \cite{disentangle_for_accent, melechovsky2024accent, melechovsky2024accented}, but require additional model training and substantial speaker-specific data.

\section{Universal Speech Content Factorization}

Speech Content Factorization (SCF) jointly decomposes content-aligned WavLM features from multiple speakers into a shared low-rank representation encoding phonetic content, enabling high-quality VC without additional model training. However, SCF is a closed-set method, as both content extraction and reconstruction require speaker-specific transformation matrices obtained during the original factorization. 
In this section, we extend SCF to an open-set setting.



\subsection{Closed-Set SCF} \label{sec:scf}

VC through SCF, as proposed in \cite{linearvc}, is performed as follows:

\begin{enumerate}
    \item Select $k$ speakers, each with sufficient speech for k-nearest-neighbor (kNN) matching (e.g., at least $5$ minutes). 
    \item For some anchor speaker $i \in [1..k]$ and associated speech WavLM frames $\mathbf{X}_i$ of shape $n$ (number of speech frames) by $d$ (WavLM feature dimension), find matching WavLM frames $\mathbf{X}_j$ using nearest neighbors for each speaker $j \in [1..k], j \neq i$. Stack the matrices $\mathbf{X}_1$ through $\mathbf{X}_k$ along their feature dimensions to produce $\mathbf{X}$ of shape $(n, kd)$.
    \item Perform rank-$r$-truncated Singular Value Decomposition (SVD) on $\mathbf{X}$, giving $\mathbf{X} \approx \mathbf{U} \mathbf{\Sigma} \mathbf{S}$. We write $\mathbf{C} = \mathbf{U} \mathbf{\Sigma}$ of shape $(n, r)$ as the content-factorized representation of $\mathbf{X}$. We further split $\mathbf{S}$ of shape $(r, kd)$ into chunks $\mathbf{S}_1$ through $\mathbf{S}_k$, each of size $(r, d)$, such that $\forall j \in [1..k]$, $\mathbf{X}_j \approx \mathbf{C} \mathbf{S}_j$. It follows that $\mathbf{X}_j \mathbf{S}_j^\dagger \approx \mathbf{C}$, where $\mathbf{S}_j^\dagger$ is the Moore-Penrose inverse of $\mathbf{S}_j$. We refer to $\mathbf{S}_j$ as the \textit{speaker transformation matrix} of speaker $j$. The content factorization process is shown in figure \ref{fig:scf}.
    \item For speakers $s, t \in [1..k]$ and input speech $\mathbf{X}'_s$, perform VC by $\hat{\mathbf{X}'}_{t} \approx \mathbf{X'}_{s}\mathbf{S}_{s}^{\dagger}\mathbf{S}_{t}$.
\end{enumerate}


\subsection{Approaches for Universal Speech to Content Mapping} \label{sec:s2c}

To extend SCF to unseen speakers, we seek:
\begin{enumerate}
    \item A matrix $\mathbf{W}$, such that for any speaker $m$ and unseen speech $\mathbf{X}_m'$, including those speakers $m$ not in $[1..k]$, $\mathbf{X}_m' \mathbf{W} \approx \mathbf{C}'$.
    \item For a small number of WavLM frames $\mathbf{X}_m'$ from an unseen speaker $m$, derive the speaker transformation matrix $\mathbf{S}_m$ corresponding to speaker $m$.
\end{enumerate}

Recall the truncated SVD formulation $\mathbf{X} \approx \mathbf{U} \mathbf{\Sigma} \mathbf{S}$.
We can solve the least-squares optimization problem $\mathbf{W}_0 = \arg\min_W \sum_{j=1}^k||\mathbf{X}_j \mathbf{W} - \mathbf{C}||$, which directly attempts to reconstruct the content-factorized representation $\mathbf{C} = \mathbf{U} \mathbf{\Sigma}$, to find $\mathbf{W}_0$. However, in preliminary experiments, we found that $\mathbf{W}_0$ does not produce representations that can be reliably mapped back to the WavLM feature space to synthesize intelligible speech. We observe that the $r$ orthonormal columns of the $n$ by $r$ matrix $\mathbf{U}$ represent the $r$ dimensions of speech content according to this factorization, the magnitude of each of these dimensions in the WavLM feature space is represented by the corresponding diagonal singular values of $\mathbf{\Sigma}$.  If we assume that all $r$ content dimensions should be treated as equally important, independent of their singular values, then $\mathbf{\Sigma}$ should be factored out of the optimization target, as shown in figure \ref{fig:ws}. This leads to the modified optimization problem on $\mathbf{U}$:

\vspace{-5mm}
\begin{align}
    \mathbf{W}_1 &=  \arg\min_W \sum_{j=1}^k||\mathbf{X}_j \mathbf{W} \mathbf{\Sigma}^{-1} - \mathbf{U}|| \label{eq:w1}
\end{align}



Using $\mathbf{U}$ directly as the optimization target may amplify acoustic contexts that are over-represented in the data. To avoid this, we instead find a matrix $\mathbf{W}_2$ that approximately inverts the speaker transformations themselves:

\vspace{-5mm}
\begin{align}
    \mathbf{W}_2 &= \arg\min_W \sum_{j=1}^k||\mathbf{S}_j \mathbf{W} - \mathbf{I}|| \label{eq:w2}
\end{align}

Finally, the formulation for $\mathbf{W}_3$ relies on the simplifying assumption that content and timbre components are linearly separable: $\mathbf{X}_j = \mathbf{C} (\mathbf{T}_{\text{content}} + \mathbf{S}_{\text{timbre}}^j)$, where $\mathbf{T}_{\text{content}}$ speaker-invariant, and its columns orthogonal to $\mathbf{S}_{\text{timbre}}^j$. 
Under this assumption we have:

\vspace{-5mm}
\begin{align}
    \mathbf{S}_i \mathbf{S}_j^\dagger &= (\mathbf{T}_{\text{content}} + \mathbf{S}_{\text{timbre}}^i) (\mathbf{T}_{\text{content}} + \mathbf{S}_{\text{timbre}}^j)^\dagger \\
    &= (\mathbf{T}_{\text{content}} + \mathbf{S}_{\text{timbre}}^i)((\mathbf{T}_{\text{content}})^\dagger + (\mathbf{S}_{\text{timbre}}^j)^\dagger) \label{eq:pseudoinverse} \\
    &= \mathbf{T}_{\text{content}} (\mathbf{T}_{\text{content}})^\dagger + \mathbf{S}_{\text{timbre}}^i (\mathbf{S}_{\text{timbre}}^j)^\dagger \label{eq:pseudoinverse_2} \\
    &\approx \mathbf{T}_{\text{content}} (\mathbf{T}_{\text{content}})^\dagger \label{eq:pseudoinverse_3} = \mathbf{I}
\end{align}

Steps \ref{eq:pseudoinverse} and \ref{eq:pseudoinverse_2} follow from the assumption of orthogonality of the column spaces of $\mathbf{T}_{\text{content}}$ and $\mathbf{S}_{\text{timbre}}^j$, while step \ref{eq:pseudoinverse_3} uses the fact that the timbre subspaces of different speakers, being high-dimensional, are likely orthogonal. Thus we can pick $\mathbf{W}_3$ simply as the Moore-Penrose inverse of any speaker $i$ in $[1..k]$:

\vspace{-5mm}
\begin{align*}
    \mathbf{W}_3 &= \mathbf{S}_i^\dagger \text{ for some } i \in [1..k]
\end{align*}

\subsection{Speaker Transformation Matrix Derivation}

Suppose we have a speaker $m \notin [1..k]$. To reconstruct speaker-$m$ speech from a content representation $\mathbf{C}$, we require the corresponding speaker transformation matrix $\mathbf{S}_m$ such that $\mathbf{X}_m \approx \mathbf{C}\mathbf{S}_m$. Suppose also that we are given a small set of WavLM features $\mathbf{X}'_m$ from speaker $m$. Using the facts that $\mathbf{X}'_m \approx \mathbf{C}' \mathbf{S}_m$ and $\mathbf{C}' \approx \mathbf{X}'_m \mathbf{W}$, we can derive $\mathbf{S}_m$ using the following formulation, also shown in figure \ref{fig:ws}:

\vspace{-5mm}
\begin{align}
    \mathbf{S}_m \approx (\mathbf{X}'_m \mathbf{W})^\dagger \mathbf{X}'_m \label{eq:std}
\end{align}



\section{Experimental setup}

\subsection{Test Data} \label{sec:data}

For our VC experiments, we select $4$ non-overlapping sets of $20$ speakers from LibriSpeech \cite{librispeech}: source speakers (from test-clean), target speakers (from test-other), held-out 1 (from test-clean), and held-out 2 (from dev-clean). For each of the source speakers, we randomly choose $20$ utterances for testing, and $5$ target speakers as the conversion target. 

\subsection{USCF-VC Details}

\begin{table}[t]
\centering
\caption{Comparisons of USCF VC with baseline systems according to objective metrics. Color intensity reflects metric directionality.}
\begin{tabular}{lccc}
\hline
Method & WER (\%) $\downarrow$ & UTMOS $\uparrow$ & Spk Sim $\uparrow$ \\
\hline
USCF $\mathbf{W}_1$ 
& \reversegradientcell{2.70}{2.18}{6.97}{white}{yellow}{50}
& \gradientcell{2.805}{2.5}{3.173}{white}{red}{50} 
& \gradientcell{0.524}{0.45}{0.666}{white}{green}{50}
\\

USCF $\mathbf{W}_2$ 
& \reversegradientcell{4.04}{2.18}{6.97}{white}{yellow}{50}
& \gradientcell{2.519}{2.5}{3.173}{white}{red}{50} 
& \gradientcell{0.557}{0.45}{0.666}{white}{green}{50}
\\

USCF $\mathbf{W}_3$ 
& \reversegradientcell{2.31}{2.18}{6.97}{white}{yellow}{50}
& \gradientcell{2.826}{2.5}{3.173}{white}{red}{50} 
& \gradientcell{0.420}{0.45}{0.666}{white}{green}{50}
\\

\hline

kNN-VC 
& \reversegradientcell{3.16}{2.18}{6.97}{white}{yellow}{50}
& \gradientcell{2.855}{2.5}{3.173}{white}{red}{50} 
& \gradientcell{0.666}{0.45}{0.666}{white}{green}{50}
\\

LinearVC 
& \reversegradientcell{2.69}{2.18}{6.97}{white}{yellow}{50}
& \gradientcell{2.765}{2.5}{3.173}{white}{red}{50} 
& \gradientcell{0.621}{0.45}{0.666}{white}{green}{50}
\\

\hline

SCF  
& \reversegradientcell{2.18}{2.18}{6.97}{white}{yellow}{50} 
& \gradientcell{2.886}{2.5}{3.173}{white}{red}{50} 
& \gradientcell{0.603}{0.45}{0.666}{white}{green}{50}
\\

SCF $\mathbf{W}_1$ 
& \reversegradientcell{3.01}{2.18}{6.97}{white}{yellow}{50}
& \gradientcell{2.859}{2.5}{3.173}{white}{red}{50} 
& \gradientcell{0.604}{0.45}{0.666}{white}{green}{50}
\\

\hline

SeedVC 
& \reversegradientcell{6.24}{2.18}{6.97}{white}{yellow}{50}
& \gradientcell{3.173}{2.5}{3.173}{white}{red}{50} 
& \gradientcell{0.532}{0.45}{0.666}{white}{green}{50}
\\

\hline
\end{tabular}
\label{tab:vc_results}
\end{table}

\begin{table}[t]
\centering
\caption{Comparison of subjective metrics between USCF and baselines. Mean scores with $95\%$ intervals.}
\begin{tabular}{lcc}
\hline
Method & MOS (\%) $\uparrow$ & SMOS (\%) $\uparrow$ \\
\hline
original 
& \gradientcell{3.80}{3.36}{3.8}{white}{red}{50} $\pm$ 0.17 
& \gradientcell{4.29}{2.77}{4.29}{white}{green}{50} $\pm$ 0.15 \\
\hline 
USCF $\mathbf{W}_1$ 
& \gradientcell{3.42}{3.36}{3.8}{white}{red}{50} $\pm$ 0.17 
& \gradientcell{3.00}{2.77}{4.29}{white}{green}{50} $\pm$ 0.23 \\
USCF $\mathbf{W}_1$ 100s 
& \gradientcell{3.66}{3.36}{3.8}{white}{red}{50} $\pm$ 0.17 
& \gradientcell{2.94}{2.77}{4.29}{white}{green}{50} $\pm$ 0.29 \\
\hline
kNN-VC 
& \gradientcell{3.53}{3.36}{3.8}{white}{red}{50} $\pm$ 0.18 
& \gradientcell{3.29}{2.77}{4.29}{white}{green}{50} $\pm$ 0.25 \\
LinearVC 
& \gradientcell{3.65}{3.36}{3.8}{white}{red}{50} $\pm$ 0.16 
& \gradientcell{3.17}{2.77}{4.29}{white}{green}{50} $\pm$ 0.21 \\
SCF 
& \gradientcell{3.63}{3.36}{3.8}{white}{red}{50} $\pm$ 0.18 
& \gradientcell{3.08}{2.77}{4.29}{white}{green}{50} $\pm$ 0.25 \\
SeedVC 
& \gradientcell{3.36}{3.36}{3.8}{white}{red}{50} $\pm$ 0.18 
& \gradientcell{2.77}{2.77}{4.29}{white}{green}{50} $\pm$ 0.31 \\
\hline
\end{tabular}
\label{tab:vc_human_results}
\end{table}

Following the procedure outlined in section \ref{sec:scf}, we choose $1$ speaker in dev-clean as our anchor speaker. The $40$ speakers included in the SVD process are drawn from held-out 1 and held-out 2. After constructing the content-aligned WavLM matrix $\mathbf{X}$, we choose $r = 75$ and perform rank-$r$-truncated SVD on $\mathbf{X}$, giving us $\mathbf{W}_1$ and $\mathbf{W}_2$ by following equations \ref{eq:w1} and \ref{eq:w2}. We set $\mathbf{W}_3$ as the pseudoinverse of $S_i$ corresponding to a different random speaker $i$ across $10$ runs, and report average performance. For each target speaker, we randomly sample from their utterances and retain the WavLM feature frames extracted from each sample utterance until we reach $500$ frames (corresponding to $10$ seconds of speech). We then derive their speaker transformation matrix according to equation \ref{eq:std}. 

\subsection{Baselines}

We include the following baselines: kNN-VC \cite{knnvc}, LinearVC \cite{linearvc}, and closed-set SCF, as these can be seen as different methods of exploiting the unique structure of the WavLM feature space. We also benchmark it against SeedVC \cite{seedvc}, a diffusion transformer-based zero-shot VC method with state-of-the-art performance. 
For kNN-VC and LinearVC, we make all source and target speaker speech (roughly $8$ minutes per speaker) available for kNN matching and for finding source-target WavLM feature mappings. 
For SCF, we perform our SVD on all $40$ speakers from the source and target speaker sets from section \ref{sec:data}, giving us $\mathbf{S}_i$ for each speaker $i$ among the source and target speakers. 

\subsection{Metrics}
We include the following objective metrics: ASR WER (using Whisper \cite{whisper} large\footnote{\scriptsize\url{huggingface.co/openai/whisper-large}}) for intelligibility; UTMOS-v2 \cite{utmos}\footnote{\scriptsize\url{huggingface.co/sarulab-speech/UTMOSv2}} as a proxy for quality; speaker embedding cosine similarity to an averaged ground-truth speaker embedding, with embeddings extracted from a pre-trained ECAPA-TDNN \cite{ecapa-tdnn}\footnote{\scriptsize\url{huggingface.co/speechbrain/spkrec-ecapa-voxceleb}}. Target Equal Error Rate, where a speaker ID system tries differentiate ground-truth targe speaker utterances from voice-converted target speaker utterances, are not reported due to space constraints, as they show identical relative performance across systems to speaker embedding cosine similarity.
We further run human evaluations on USCF and comparative systems, where we collect a Mean Opinion Score (MOS) based on the naturalness and quality of each output utterance, and a Speaker Similarity Mean Opinion Score (SMOS) based on the similarity of each output utterance to a reference utterance.

\section{Results}
\subsection{Voice Conversion Quality}

Objective evaluation results for voice conversion are presented in table \ref{tab:vc_results}. We note that USCF excel at content preservation, and is capable of producing natural-sounding speech. Speaker similarity tests show that while USCF is able to produce speech that is highly similar to the target speaker, the level of similarity is somewhat weaker than those of kNN-VC, LinearVC, and SCF. In order to investigate the source of this degradation, we compare USCF with partially open-set SCF, where the input speech comes from speakers that are out-of-domain while the target speakers are in-domain. We find that partially open-set SCF achieves speaker similarities that are on par with closed-set SCF, suggesting that the content-to-speaker transformation is the source of the degradation in speaker similarity of USCF. 

Subjective evaluation results, shown in table \ref{tab:vc_human_results}, demonstrate that listeners show no statistically significant preference towards USCF or any of the other baseline systems except SeedVC, which was least favored. 

Among the speech-to-content mapping strategies, we found that $\mathbf{W}_2$ achieves the best target speaker similarity at the price of reduced speech quality and content preservation, whereas $\mathbf{W}_3$ shines at content preservation but struggles with target speaker similarity; by contrast, $\mathbf{W}_1$ finds a reasonable balance between all metrics. For USCF using $\mathbf{W}_3$ as the universal speech-to-content mapping, we report the average performance over $10$ runs, where $\mathbf{W}_3$ was chosen to be the pseudoinverse of a different randomly-selected SCF speaker transformation matrix. Over the $10$ runs, the standard deviation for ASR WER and UTMOS are $0.10\%$ and $0.015$, respectively, suggesting that $\mathbf{W}_3$ is highly stable regardless of the choice of SCF speaker transformation used to derive it.

\subsection{Speaker ID within Phoneme}
\begin{table}[t]
\centering
\caption{Speaker ID using same-phoneme embeddings.}
\begin{tabular}{lccc}
\hline
Model & Rank & Spk EER $\uparrow$ & Phoneme EER $\downarrow$ \\
\hline
USCF  & 75   & \textbf{36.40\%}  & 11.43\% \\
USCF  & 1024 & 35.33\% & 11.43\% \\
\hline
WavLM & 1024 & 21.77\% & 11.56\% \\
ContentVec & 792 & 27.98\% & \textbf{8.82\%} \\ 
WavLM - kNN & 1024 & 35.14\% & 12.70\% \\
\hline
\end{tabular}
\label{tab:spk_eer}
\end{table}

\begin{table}[t]
\centering
\caption{VC performance of USCF across different ranks.}
\begin{tabular}{lccc}
\hline
Rank & WER (\%) $\downarrow$ & UTMOS $\uparrow$ & Spk Sim $\uparrow$ \\
\hline
100 
& \reversegradientcell{2.77}{2.18}{6.97}{white}{yellow}{50} 
& \gradientcell{2.81}{2.5}{3.173}{white}{red}{50} 
& \gradientcell{0.504}{0.45}{0.666}{white}{green}{50} 
\\

75  
& \reversegradientcell{2.70}{2.18}{6.97}{white}{yellow}{50} 
& \gradientcell{2.805}{2.5}{3.173}{white}{red}{50} 
& \gradientcell{0.524}{0.45}{0.666}{white}{green}{50} 
\\

50  
& \reversegradientcell{2.69}{2.18}{6.97}{white}{yellow}{50} 
& \gradientcell{2.738}{2.5}{3.173}{white}{red}{50} 
& \gradientcell{0.529}{0.45}{0.666}{white}{green}{50} 
\\

30  
& \reversegradientcell{2.96}{2.18}{6.97}{white}{yellow}{50} 
& \gradientcell{2.607}{2.5}{3.173}{white}{red}{50} 
& \gradientcell{0.513}{0.45}{0.666}{white}{green}{50} 
\\

20  
& \reversegradientcell{3.98}{2.18}{6.97}{white}{yellow}{50} 
& \gradientcell{2.388}{2.5}{3.173}{white}{red}{50} 
& \gradientcell{0.489}{0.45}{0.666}{white}{green}{50} 
\\

\hline
\end{tabular}
\label{tab:rank_results}
\end{table}

\begin{table}[t]
\centering
\caption{VC performance of USCF when output output speaker transformation matrix is derived with different number of frames of target speaker speech.}
\begin{tabular}{lcccc}
\hline
Num frames & WER (\%) $\downarrow$ & UTMOS $\uparrow$ & Spk Sim $\uparrow$ \\
\hline
10000 
& \reversegradientcell{2.28}{2.18}{6.97}{white}{yellow}{50} 
& \gradientcell{2.935}{2.5}{3.173}{white}{red}{50}  
& \gradientcell{0.564}{0.45}{0.666}{white}{green}{50} 
\\

5000 
& \reversegradientcell{2.42}{2.18}{6.97}{white}{yellow}{50} 
& \gradientcell{2.923}{2.5}{3.173}{white}{red}{50}  
& \gradientcell{0.564}{0.45}{0.666}{white}{green}{50}
\\

2000 
& \reversegradientcell{2.47}{2.18}{6.97}{white}{yellow}{50} 
& \gradientcell{2.915}{2.5}{3.173}{white}{red}{50}  
& \gradientcell{0.564}{0.45}{0.666}{white}{green}{50} 
\\

1000 
& \reversegradientcell{2.51}{2.18}{6.97}{white}{yellow}{50} 
& \gradientcell{2.904}{2.5}{3.173}{white}{red}{50} 
& \gradientcell{0.546}{0.45}{0.666}{white}{green}{50} 
\\

500 
& \reversegradientcell{2.70}{2.18}{6.97}{white}{yellow}{50} 
& \gradientcell{2.805}{2.5}{3.173}{white}{red}{50} 
& \gradientcell{0.524}{0.45}{0.666}{white}{green}{50} 
\\

200 
& \reversegradientcell{4.94}{2.18}{6.97}{white}{yellow}{50} 
& \gradientcell{2.431}{2.5}{3.173}{white}{red}{50} 
& \gradientcell{0.42}{0.45}{0.666}{white}{green}{50} 
\\

\hline
\end{tabular}
\label{tab:frame_limits}
\end{table}
\begin{table}[t]
\centering
\footnotesize
\caption{TTS models trained using different target features. Fbank (normalized) is where each utterance is first normalized to the same voice using kNN-VC \cite{knnvc}, then converted into Fbank features.}
\begin{tabular}{lccc}
\hline
Target Features & ASR WER $\downarrow$ & Epochs & UTMOS-v2 $\uparrow$ \\
\hline
USCF features & 11.44\% & 25 & 2.881 \\
\hline
mel & 27.93\% & 39 & 2.741 \\
mel (normalized) & 11.92\% & 33 & 2.732 \\
\hline
\end{tabular}
\label{tab:tts}
\end{table}
\vspace{-3mm}

\vspace{5mm}
To test the claim that USCF features preserve speech content information while removing speaker-identifying information, we first extract USCF features (rank $75$, $\mathbf{W}_1$) from the TIMIT \cite{timit} dataset TEST split, then perform the following:
\begin{enumerate}
    \item Phoneme recognition. Ground truth phoneme labels for each test frame are inferred using TIMIT's time-phoneme alignment. Classification is performed by taking the argmin cosine distance to each reference phoneme embedding.
    \item Speaker recognition per-phoneme. For each phoneme, we find all the frames from TIMIT TEST with the corresponding phoneme label, then perform speaker ID on feature vectors corresponding to these frames.
\end{enumerate}

We compare USCF to two SSL-based baselines: WavLM and ContentVec, shown in table \ref{tab:spk_eer}. We found that USCF is on par with WavLM as a phoneme classifier, and that it is stronger than both WavLM and ContentVec when it comes to removing speaker information. This property still holds even when the rank of USCF is increased to $1024$, showing that the loss of speaker information is not an artifact of projecting into a low-dimensional space. We additionally find that USCF is still marginally better at speaker information removal and at content preservation than WavLM features that had been normalized to a single reference speaker through kNN matching.

\subsection{Ablations}

\subsubsection{Reduction in Rank}

A comparison of the performance of USCF at various ranks can be found in table \ref{tab:rank_results}. We find that USCF is stable when its rank is within $50$ and $100$, but that the synthetic voice quality declines when its rank drops further. 

\subsubsection{$\mathbf{S}$ Derivation with Fewer or More Frames}

Table \ref{tab:frame_limits} shows the performance of USCF when the amount of target speaker speech used for deriving the target speaker transformation matrix is varied. We notice a sharp degradation in target speaker similarity when the amount of target speaker speech falls below $500$ frames, corresponding to $10$ seconds; conversely, target speaker similarity improves when the amount of target speaker speech is increased, but see diminishing returns beyond $2000$ frames ($40$ seconds).

\subsection{TTS Using a Content-Factorized Representation}

We verify that USCF features can be used as target training features with a flow-matching TTS model, trained on LibriSpeech and following the architecture and training strategies from ZipVoice \cite{zipvoice}.
We compare it to two benchmarks: a TTS model trained on mel filterbank features, and one trained on mel filterbank features extracted from utterances that had been voice-normalized to a single speaker using kNN-VC. Table \ref{tab:tts} show that the model trained with USCF features achieves better WER while requiring less training time than both models trained with mel filterbank features. 

\section{Conclusion}

We propose USCF, a method for extracting content-preserving and speaker-agnostic features from WavLM features using a simple linear transformation. We extend an existing closed-set method for linear factorization of WavLM features, SCF, into an open-set method, transforming it into a zero-shot voice conversion system. We show through embedding analysis that USCF features effectively preserve content information, and that they carry less speaker information than existing methods such as ContentVec. We also demonstrate the potential for using USCF features on downstream tasks such as TTS training. In our future work, we would like to explore if simple neural methods would allow for a more stable version of $\mathbf{W}$, or a derivation of $\mathbf{S}_m$ for unseen speaker $m$ with even less speech from speaker $m$ available. We also plan to utilize USCF to train zero-shot style-conditioned TTS systems that are timbre-agnostic.

\section{Generative AI Disclosure}

Generative AI was used for the following in this work:

\begin{enumerate}
    \item Conversion of table formats.
    \item Limited language polishing of the Introduction section.
\end{enumerate}

\ifcameraready
\section{Acknowledgments}
     This work was supported by the Office of the Director of National Intelligence (ODNI), Intelligence Advanced Research Projects Activity (IARPA), via the ARTS Program under contract D2023-2308110001. The views and conclusions contained herein are those of the authors and should not be interpreted as necessarily representing the official policies, either expressed or implied, of ODNI, IARPA, or the U.S. Government. The U.S. Government is authorized to reproduce and distribute reprints for governmental purposes notwithstanding any copyright annotation therein.
\else
\fi

\bibliographystyle{IEEEtran}
\bibliography{mybib}

@inproceedings{linearvc,
  title     = {{LinearVC: Linear Transformations of Self-Supervised Features Through the Lens of Voice Conversion}},
  author    = {Herman Kamper and Benjamin {van Niekerk} and Julian Zaïdi and Marc-André Carbonneau},
  year      = {2025},
  booktitle = {{Interspeech 2025}},
  pages     = {1398--1402},
  doi       = {10.21437/Interspeech.2025-438},
  issn      = {2958-1796},
}

@inproceedings{melechovsky2024accent,
  title={Accent conversion in text-to-speech using multi-level VAE and adversarial training},
  author={Melechovsky, Jan and Mehrish, Ambuj and Sisman, Berrak and Herremans, Dorien},
  booktitle={TENCON 2024-2024 IEEE Region 10 Conference (TENCON)},
  pages={473--476},
  year={2024},
  organization={IEEE}
}

@inproceedings{melechovsky2024accented,
  title={Accented text-to-speech synthesis with a conditional variational autoencoder},
  author={Melechovsky, Jan and Mehrish, Ambuj and Sisman, Berrak and Herremans, Dorien},
  booktitle={TENCON 2024-2024 IEEE Region 10 Conference (TENCON)},
  pages={343--346},
  year={2024},
  organization={IEEE}
}

@article{sisman2020overview,
  title={An overview of voice conversion and its challenges: From statistical modeling to deep learning},
  author={Sisman, Berrak and Yamagishi, Junichi and King, Simon and Li, Haizhou},
  journal={IEEE/ACM transactions on audio, speech, and language processing},
  volume={29},
  pages={132--157},
  year={2020},
  publisher={IEEE}
}

@inproceedings{knnvc,
  author={Matthew Baas and Benjamin {van Niekerk} and Herman Kamper},
  title={{Voice Conversion With Just Nearest Neighbors}},
  booktitle={Interspeech 2023},
  pages={2053--2057},
}

@misc{vae,
      title={Auto-Encoding Variational Bayes}, 
      author={Diederik P Kingma and Max Welling},
      year={2022},
      eprint={1312.6114},
      archivePrefix={arXiv},
      primaryClass={stat.ML},
      url={https://arxiv.org/abs/1312.6114}, 
}

@misc{zipvoice,
      title={ZipVoice: Fast and High-Quality Zero-Shot Text-to-Speech with Flow Matching}, 
      author={Han Zhu and Wei Kang and Zengwei Yao and Liyong Guo and Fangjun Kuang and Zhaoqing Li and Weiji Zhuang and Long Lin and Daniel Povey},
      year={2025},
      eprint={2506.13053},
      archivePrefix={arXiv},
      primaryClass={eess.AS},
      url={https://arxiv.org/abs/2506.13053}, 
}

@inproceedings{wavlm_phoneme_cluster,
    title = "Leveraging Allophony in Self-Supervised Speech Models for Atypical Pronunciation Assessment",
    author = "Choi, Kwanghee  and
      Yeo, Eunjung  and
      Chang, Kalvin  and
      Watanabe, Shinji  and
      Mortensen, David R",
    booktitle = "Proceedings of the 2025 Conference of the Nations of the Americas Chapter of the Association for Computational Linguistics: Human Language Technologies (Volume 1: Long Papers)",
    year = "2025",
    pages = "2613--2628",
}

@inproceedings{wavlm_phoneme_recognition,
  title     = {{Self-Supervised Models for Phoneme Recognition: Applications in Children's Speech for Reading Learning}},
  author    = {Lucas {Block Medin} and Thomas Pellegrini and Lucile Gelin},
  year      = {2024},
  booktitle = {{Interspeech 2024}},
  pages     = {5168--5172},
  doi       = {10.21437/Interspeech.2024-1095},
  issn      = {2958-1796},
}

@misc{genvc,
      title={GenVC: Self-Supervised Zero-Shot Voice Conversion}, 
      author={Zexin Cai and Henry Li Xinyuan and Ashi Garg and Leibny Paola García-Perera and Kevin Duh and Sanjeev Khudanpur and Matthew Wiesner and Nicholas Andrews},
      year={2025},
      eprint={2502.04519},
      archivePrefix={arXiv},
      primaryClass={eess.AS},
      url={https://arxiv.org/abs/2502.04519}, 
}

@misc{seedvc,
      title={Zero-shot Voice Conversion with Diffusion Transformers}, 
      author={Songting Liu},
      year={2024},
      eprint={2411.09943},
      archivePrefix={arXiv},
      primaryClass={cs.SD},
      url={https://arxiv.org/abs/2411.09943}, 
}

@inproceedings{disentangle_for_vc,
  title     = {{Automatic Speech Disentanglement for Voice Conversion using Rank Module and Speech Augmentation}},
  author    = {Zhonghua Liu and Shijun Wang and Ning Chen},
  year      = {2023},
  booktitle = {{Interspeech 2023}},
  pages     = {2298--2302},
  doi       = {10.21437/Interspeech.2023-1602},
  issn      = {2958-1796},
}

@inproceedings{disentangle_for_st,
    title = "{CCSRD}: Content-Centric Speech Representation Disentanglement Learning for End-to-End Speech Translation",
    author = "Zhao, Xiaohu  and
      Sun, Haoran  and
      Lei, Yikun  and
      Zhu, Shaolin  and
      Xiong, Deyi",
    booktitle = "Findings of the Association for Computational Linguistics: EMNLP 2023",
    year = "2023",
    pages = "5920--5932",
}

@misc{disentangle_for_privacy,
      title={Universal Semantic Disentangled Privacy-preserving Speech Representation Learning}, 
      author={Biel Tura Vecino and Subhadeep Maji and Aravind Varier and Antonio Bonafonte and Ivan Valles and Michael Owen and Leif Rädel and Grant Strimel and Seyi Feyisetan and Roberto Barra Chicote and Ariya Rastrow and Constantinos Papayiannis and Volker Leutnant and Trevor Wood},
      year={2025},
      eprint={2505.13085},
      archivePrefix={arXiv},
      primaryClass={eess.AS},
      url={https://arxiv.org/abs/2505.13085}, 
}

@misc{disentangle_for_accent,
      title={FAC-FACodec: Controllable Zero-Shot Foreign Accent Conversion with Factorized Speech Codec}, 
      author={Yurii Halychanskyi and Cameron Churchwell and Yutong Wen and Volodymyr Kindratenko},
      year={2026},
      eprint={2510.10785},
      archivePrefix={arXiv},
      primaryClass={cs.SD},
      url={https://arxiv.org/abs/2510.10785}, 
}

@inproceedings{npu_ntu,
  title     = {{NPU-NTU System for Voice Privacy 2024 Challenge}},
  author    = {Jixun Yao and Nikita Kuzmin and Qing Wang and Pengcheng Guo and Ziqian Ning and Dake Guo and Kong Aik Lee and Eng-Siong Chng and Lei Xie},
  year      = {2024},
  booktitle = {{4th Symposium on Security and Privacy in Speech Communication}},
  pages     = {67--71},
  doi       = {10.21437/SPSC.2024-12},
}

@inproceedings{naturalspeech3,
  author = {Ju, Zeqian and Wang, Yuancheng and Shen, Kai and Tan, Xu and Xin, Detai and Yang, Dongchao and Liu, Eric and Leng, Yichong and Song, Kaitao and Tang, Siliang and Wu, Zhizheng and Qin, Tao and Li, Xiangyang and Ye, Wei and Zhang, Shikun and Bian, Jiang and He, Lei and Li, Jinyu and Zhao, Sheng},
  booktitle = {ICML},
  pages = {22605-22623},
  title = {NaturalSpeech 3: Zero-Shot Speech Synthesis with Factorized Codec and Diffusion Models.},
  volume = 235,
  year = 2024
}

@inproceedings{continuous_codec,
  author = {Xie, Yuying and Kuhlmann, Michael and Rautenberg, Frederik and Tan, Zheng-Hua and Haeb-Umbach, Reinhold},
  booktitle = {EUSIPCO},
  pages = {436-440},
  publisher = {IEEE},
  title = {Speaker and Style Disentanglement of Speech Based on Contrastive Predictive Coding Supported Factorized Variational Autoencoder.},
  year = 2024
}

@misc{speechsplit2,
      title={SpeechSplit 2.0: Unsupervised speech disentanglement for voice conversion Without tuning autoencoder Bottlenecks}, 
      author={Chak Ho Chan and Kaizhi Qian and Yang Zhang and Mark Hasegawa-Johnson},
      year={2022},
      eprint={2203.14156},
      archivePrefix={arXiv},
      primaryClass={eess.AS},
      url={https://arxiv.org/abs/2203.14156}, 
}

@inproceedings{adv_disentangled,
  title     = {{Adversarially Learning Disentangled Speech Representations for Robust Multi-Factor Voice Conversion}},
  author    = {Jie Wang and Jingbei Li and Xintao Zhao and Zhiyong Wu and Shiyin Kang and Helen Meng},
  year      = {2021},
  booktitle = {{Interspeech 2021}},
  pages     = {846--850},
  doi       = {10.21437/Interspeech.2021-1990},
  issn      = {2958-1796},
}

@InProceedings{contentvec,
  title = 	 {{C}ontent{V}ec: An Improved Self-Supervised Speech Representation by Disentangling Speakers},
  author =       {Qian, Kaizhi and Zhang, Yang and Gao, Heting and Ni, Junrui and Lai, Cheng-I and Cox, David and Hasegawa-Johnson, Mark and Chang, Shiyu},
  booktitle = 	 {Proceedings of the 39th International Conference on Machine Learning},
  pages = 	 {18003--18017},
  year = 	 {2022},
  volume = 	 {162},
  series = 	 {Proceedings of Machine Learning Research},
}

@inproceedings{dvae,
  title     = {{Speech Resynthesis from Discrete Disentangled Self-Supervised Representations}},
  author    = {Adam Polyak and Yossi Adi and Jade Copet and Eugene Kharitonov and Kushal Lakhotia and Wei-Ning Hsu and Abdelrahman Mohamed and Emmanuel Dupoux},
  year      = {2021},
  booktitle = {{Interspeech 2021}},
  pages     = {3615--3619},
  doi       = {10.21437/Interspeech.2021-475},
  issn      = {2958-1796},
}

@inproceedings{ecapa-tdnn,
  title     = {{ECAPA-TDNN: Emphasized Channel Attention, Propagation and Aggregation in TDNN Based Speaker Verification}},
  author    = {Brecht Desplanques and Jenthe Thienpondt and Kris Demuynck},
  year      = {2020},
  booktitle = {{Interspeech 2020}},
  pages     = {3830--3834},
  doi       = {10.21437/Interspeech.2020-2650},
  issn      = {2958-1796},
}

@misc{whisper,
      title={Robust Speech Recognition via Large-Scale Weak Supervision}, 
      author={Alec Radford and Jong Wook Kim and Tao Xu and Greg Brockman and Christine McLeavey and Ilya Sutskever},
      year={2022},
      eprint={2212.04356},
      archivePrefix={arXiv},
      primaryClass={eess.AS},
      url={https://arxiv.org/abs/2212.04356}, 
}

@inproceedings{utmos,
  title     = {{UTMOS: UTokyo-SaruLab System for VoiceMOS Challenge 2022}},
  author    = {Takaaki Saeki and Detai Xin and Wataru Nakata and Tomoki Koriyama and Shinnosuke Takamichi and Hiroshi Saruwatari},
  year      = {2022},
  booktitle = {{Interspeech 2022}},
  pages     = {4521--4525},
  doi       = {10.21437/Interspeech.2022-439},
  issn      = {2958-1796},
}

@inproceedings{commonvoice,
    title={{Common Voice: A Massively-Multilingual Speech Corpus}},
  author = "Ardila, Rosana  and
      Branson, Megan  and
      Davis, Kelly  and
      Kohler, Michael  and
      Meyer, Josh  and
      Henretty, Michael  and
      Morais, Reuben  and
      Saunders, Lindsay  and
      Tyers, Francis  and
      Weber, Gregor",
  booktitle = "Proceedings of the Twelfth Language Resources and Evaluation Conference",
  pages = "4218--4222",
  year={2020}
}

@INPROCEEDINGS{emilia,
  author={He, Haorui and Shang, Zengqiang and Wang, Chaoren and Li, Xuyuan and Gu, Yicheng and Hua, Hua and Liu, Liwei and Yang, Chen and Li, Jiaqi and Shi, Peiyang and Wang, Yuancheng and Chen, Kai and Zhang, Pengyuan and Wu, Zhizheng},
  booktitle={2024 IEEE Spoken Language Technology Workshop (SLT)}, 
  title={Emilia: An Extensive, Multilingual, and Diverse Speech Dataset For Large-Scale Speech Generation}, 
  year={2024},
  volume={},
  number={},
  pages={885-890},
  doi={10.1109/SLT61566.2024.10832365}}

@INPROCEEDINGS{librispeech,
  author={Panayotov, Vassil and Chen, Guoguo and Povey, Daniel and Khudanpur, Sanjeev},
  booktitle={2015 IEEE International Conference on Acoustics, Speech and Signal Processing (ICASSP)}, 
  title={Librispeech: An ASR corpus based on public domain audio books}, 
  year={2015},
  volume={},
  number={},
  pages={5206-5210},
  doi={10.1109/ICASSP.2015.7178964}}

@misc{timit,
  author = {Garofolo, John S. and of Standards, National Institute and U.S., Technology and States, United and Agency., Defense Advanced Research Projects and Science, Information and Office, Technology and Consortium., Linguistic Data},
  title = {{TIMIT : acoustic-phonetic continuous speech corpus.}},
  url = {http://www.worldcat.org/isbn/1585630195},
  year = 1993
}

\end{document}